# Multilevel classification framework for breast cancer cell selection and its integration with advanced disease models

Catarina Franco Jones,[1,2] Diogo Dias,[3,4] Ana C. Moreira,[5,6] Gil Gonçalves,[5,6] Stefano Cinti,[7] Mustafa B. A. Djamgoz,[8] Frederico Castelo Ferreira,[3,4] Paola Sanjuán-Alberte,[3,4] Rosalia Moreddu[1,2,*]

[1]School of Electronics and Computer Science, University of Southampton, Southampton SO171BJ, UK
[2]Institute for Life Sciences, University of Southampton, Southampton SO17 1BJ, UK
[3]Department of Bioengineering, Instituto Superior Técnico, Universidade de Lisboa, 1049-001 Lisbon, Portugal
[4]Institute for Health and Bioeconomy, Instituto Superior Técnico, Universidade de Lisboa, 1049-001 Lisbon, Portugal
[5]Centre for Mechanical Technology and Automation (TEMA), University of Aveiro, 3810-193 Aveiro, Portugal
[6]Intelligent Systems Associate Laboratory (LASI), 4800-058 Guimarães, Portugal
[7]Department of Pharmacy, University of Naples "Federico II", Naples, Italy
[8]Department of Life Sciences, Imperial College London, SW7 2BX London, UK
*Correspondence: r.moreddu@soton.ac.uk

## SUMMARY

Breast cancer cell lines are indispensable tools for unraveling disease mechanisms, enabling drug discovery, and developing personalized treatments, yet their heterogeneity and inconsistent classification pose significant challenges in model selection and data reproducibility. This review aims at providing a comprehensive and user-friendly framework for broadly mapping the features of breast cancer types and commercially available human breast cancer cell lines, defining *absolute criteria*, i.e. objective features such as origin (e.g., MDA-MB, MCF), histological subtype (ductal, lobular), hormone receptor status (ER/PR/HER2), and genetic mutations (BRCA1, TP53), and *relative criteria*, which contextualize functional behaviors like metastatic potential, drug sensitivity, and genomic instability. It then examines how the proposed framework could be applied to cell line screening in advanced and emerging disease models. By supporting better informed choices, this work aims to improve experimental design and strengthen the connection between *in vitro* breast cancer studies and their clinical translation.

## INTRODUCTION

Breast cancer cell lines are *in vitro* disease models widely used in biomedical research to gain insights into the pathophysiology of the disease, and to develop novel diagnostic and therapeutic strategies.[1] Derived from human tumors, they provide a renewable resource for investigating the cellular and molecular mechanisms underlying disease progression, enabling the search for new therapeutic agents and diagnostic markers, by recapitulating local conditions and allowing controlled perturbations *in vitro*.[2,3] In drug discovery and development, cell models are utilized to screen potential anti-cancer drugs for their efficacy and possible toxicity[4]. In personalized medicine, patient-derived cells allow to evaluate individual responses to specific treatments, aiming to improve therapeutic outcomes.[5] However, cell lines do not fully represent the heterogeneity of patient tumors, especially when employed in isolation, risking to oversimplify the biological environments characteristic of complex living systems.[6] Moreover, they may acquire genetic changes during long-term culture, leading to substantial alterations in both morphology and functionality.[6] In this scenario, selecting the optimal cancer cell line based on its properties and experimental objectives becomes critical toward obtaining reproducible and translatable results. This review presents a classification framework that distinguishes commercially available human breast cancer cell lines based on absolute criteria, such as origin and hormone receptor status, and relative criteria, such as metastatic potential and drug response. It illustrates how diverse cellular features can be systematically organized to optimize cell lines use in translational research, alongside their integration with advanced disease models, from organoids to co-culture systems and patient-derived xenografts.

Absolute criteria form the foundational layer of cell lines classification, capturing static or semi-static features that define the cell line identity and heritage. The *origin* of the cell line, representing whether it is derived from a primary tumor or metastatic lesion, is a clear example of absolute criteria, and influences differentiation state and drug responsiveness. Histological subtype, including ductal, lobular, or metaplastic, offers another level of biological context that can dictate architectural and invasive properties.[7] Hormone receptor status (ER, PR, HER2), one of the most clinically relevant stratifications in breast oncology[8], drives therapeutic decisions and aligns with subtypes such as triple-negative or HER2-positive disease. Moreover, molecular subtypes as luminal A/B or basal-like capture transcriptomic signatures that reflect biological states[9], while common genetic mutations (e.g., BRCA1/2, TP53, PTEN) frame the cell within defined oncogenic trajectories.[10] Absolute parameters ensure that the use of a cell line is grounded in clinically and genetically meaningful choices; yet, relying solely on absolute descriptors neglects the functional plasticity of breast cancer cells.

The relative criteria component addresses this limitation by encompassing dynamic phenotypes and regulatory states that vary across contexts and influence experimental outputs. These include functional attributes such as metastatic ability, proliferation rate and apoptotic resistance, which directly affect how a cell line behaves under experimental perturbation. For example, the propensity to metastasize to bone or brain, which is central to studying organotropism, is interlinked with signaling adaptations and cellular machinery, such as invadopodia formation and EMT programs. [11,12] Similarly, apoptotic resistance, shaped by alterations in caspase expression or FLIP activity, dictates the cell line survival under cytotoxic challenge and is therefore crucial for modeling drug resistance.[13] The relative aspects also incorporate advanced molecular and regulatory dimensions such as gene expression signatures (e.g., PAM50, MammaPrint), radiation response profiles, and drug sensitivity patterns that mirror therapeutic resistance observed in clinical settings. This stratification aligns with the contemporary view that cancer is a highly adaptive and evolving system.[14] Additionally, layers such as epigenomic modifications, stem cell properties, and inflammatory status bridge the molecular and microenvironmental axes of tumor biology. For instance, ALDH1 or CD44+/CD24− status has been linked to tumor-initiating potential and chemoresistance[15], and immune signatures (e.g., PD-L1 or cytokine expression) can dictate immunotherapeutic outcomes.[16] Genomic instability adds another dimension, often indicating susceptibility to specific DNA-damaging agents or synthetic lethality strategies. [17]

This schematization emphasizes how integrating these criteria could drive the choice and development of more predictive and clinically relevant models. Multi-omics could then allow simultaneous profiling of genomics, transcriptomics, and proteomics to refine cell line characterization, enabling matching to patient-derived data. AI and data fusion techniques could forecast drug responses or disease trajectories based on integrative datasets. Finally, informed baseline choices could guide the design of *in vitro* models. 3D tissue constructs and co-culture systems better recapitulate *in vivo* architectures and cell-cell interactions, serving as an essential bridge toward preclinical and patient-derived xenograft (PDX) models, ensuring that selected cell lines can perform in biologically complex environments or be employed for their development.

## ABSOLUTE CLASSIFICATION CRITERIA

We define as *absolute criteria* those features that can be classified statically and objectively. As opposed to *relative criteria* (e.g., metastatic potential graded as *low* or *high*), absolute criteria are intrinsic attributes such as cellular origin (e.g., “MCF” origin, Michigan Cancer Foundation), histological subtype, hormone receptor status, genetic alterations, and molecular subtype. The following subsections address these characteristics.

### *Cell line origin*

The origin of a cell line denotes its derivation from specific breast tumor tissues. This classification is rooted in the cell line provenance and drives its research application. For brevity, the origins of breast cancer cell line families, along with their associated experimental uses, are summarized in **Table 1**. In addition to tumor-derived breast cancer cell lines, engineered breast epithelial models have been developed to study specific processes like transformation, epithelial-to-mesenchymal transition (EMT), and cancer stem cell behavior in a more controlled context. A widely used model is MCF10A, a non-tumorigenic human mammary epithelial cell line often used as the healthy reference in breast cancer studies.[18] More advanced systems include HMLE and HMLER cells. HMLE cells are obtained by introducing genes that prevent cellular aging (hTERT) and inhibit tumor suppressor activity (SV40 large T antigen), allowing long-term growth in culture.[19] When the HMLE model is further modified with an oncogene (H-RAS), it becomes tumorigenic (known as HMLER).[19] A variant called HMLER-shEcad, where the gene for E-cadherin is silenced, is commonly used to model EMT and the acquisition of cancer stem cell–like traits.[19] Despite lacking the genetic complexity of actual tumors, these models are valuable tools to dissect the functional impact of specific molecular changes and complement the use of patient-derived breast cancer cell lines in experimental research.

**Table 1. Cell line family, acronym origin, representative cell lines, and clinically relevant research applications of cell line families.**

| Cell line family | Acronym origin | Representative cell lines | | Key research applications of the cell line family | Ref. |
|---|---|---|---|---|---|
| | | *Name* | *Source* | | |
| **MDA-MB** | M.D. Anderson Cancer Center - Mammary/Breast | MDA-MB-231 | Metastatic sites, pleural effusions | Metastasis; chemoresistance; tumor-microenvironment interactions | 20 |
| | | MDA-MB-468 | Brain metastasis | | |
| **MCF** | Michigan Cancer Foundation | MCF-7 | Pleural effusions | HR+ Breast cancer progression; weakly metastatic control | 21 |
| | | MCF-10A | Fibrocystic breast tissue | | |
| **HCC** | Human Cancer Culture | HCC1937 | Primary breast tumor carrying BRCA-1 mutation | DNA repair defects; targeted therapy resistance | 22 |
| | | HCC1954 | HER2-positive metastatic site | | |
| **BT** | Breast Tumor | BT-474 | solid invasive ductal carcinoma, HER2 amplification | HER2-targeted therapies (drug testing, e.g., lapatinib) | 23 |
| | | BT-20 | Primary TNBC, lacks functional TP53 | | |
| **CAMA** | Caucasian Malignant Adenocarcinoma | CAMA-1 | Liver metastasis | Endocrine resistance mechanisms | 24 |
| **SK-BR** | Sloan Kettering Institute – Breast | SK-BR-3 | Pleural effusion, TP53 mutation | HER2-targeted therapies | 25 |
| **ZR** | Zurich/Michigan cancer foundation | ZR-75-1 | Ascitic effusion, metastatic ductal carcinoma | Hormone receptor plasticity; metastatic adaptation | 26 |
| | | ZR-75-30 | Subline of the above, with reduced hormone dependence | | |
| **SUM** | Dr. Stephen Ethier, University of Michigan | SUM-149PT | Primary inflammatory TNBC, BRCA-1 mutation | IBC-specific pathways | 27 |
| | | SUM-159PT | Metastatic site of inflammatory TNBC | | |
| **Hs** | Human Somatic | Hs578T | Breast carcinosarcoma | Sarcomatoid differentiation; tumor-stroma crosstalk | 28 |
| **DU** | Duke University | DU4475 | Rare metastatic TNBC model | niche-specific metastasis mechanisms | 29 |
| **CAL** | Cancer Associated Line | CAL-51 | Ductal carcinoma, TP53-mutated | tumor heterogeneity | 30 |
| | | CAL-120 | Metastatic site, basal-like | | |
| **MFM** | Max Faber Memorial laboratory | MFM-223 | Pleural effusion with metaplastic TNBC | tumor-stroma interactions; drug sensitivity in metaplastic carcinomas | 31 |
| **PMC** | Primary Malignant Culture | PMC-42 | Invasive ductal carcinoma, forms organoids *in vitro* | Morphogenesis; polarization; extracellular matrix role in tumor progression | 32 |
| **UACC** | University of Arizona Cancer Center | UACC-812 | metastatic site (likely lymph node) | Drug resistance; gene signatures and drug response | 33 |
| | | UACC-893 | HER2-positive ductal carcinoma | | |
| **EMG** | Epidermal Malignant Growth | EM-G3 | scirrhous carcinoma, desmoplastic subtype | Desmoplasia; tumor microenvironment crosstalk | 34 |
| **HDQ** | *Unknown* | HDQ-P1 | Primary ductal carcinoma with BRCA-2 mutations | Synthetic lethality strategies; resistance mechanisms | [35] |
| **EFM** | European Foundation for Medicine | EFM-19 | Malignant pleural effusion | epigenomic modifications; alternative survival pathways | [36] |
| **IBEP** | Instituto de Biomedicina, Estudio de Proliferación | IBEP-1 | Invasive ductal carcinoma, luminal-like | intratumoral heterogeneity; clonal evolution | [37] |
| | | IBEP-2 | Invasive ductal carcinoma, basal-like | | |
| **KPL** | Kurebayashi Pleural Line | KPL-1 | malignant ascites of a HER2-positive patient | antibody-drug conjugate mechanisms | [38] |
| **LY** | Dr. Anne Lykkesfeldt | LY-2 | tamoxifen-resistant subline of MCF-7 | HR and growth factor pathways; role of autophagy in acquired resistance | [39] |
| **T** | Tissue culture | T-47D | Pleural effusion | progesterone receptor (PR) signaling; CDK4/6 inhibitor responses | [40] |
| **BSMZ** | Bützow, Sager, Müller, Zurich | BSMZ | Mucinous carcinoma | glycoprotein-mediated immune evasion and matrix adhesion | [18,41] |
| **AU** | Auburn University | AU565 | Metastatic site | antibody-drug conjugates | [42] |
| **21** | Age of patient (21 years old) | 21-MT-1 | Metastatic breast tumor | PARP inhibitor responses; metastasis-initiating cells | [43] |
| | | 21-PT | Primary breast tumor | | |
| **HMT** | Hanyang Medical Team | HMT-3902S1 | Primary breast tumor | TGF-β-driven EMT and metastasis in xenograft models | [44] |
| **MA** | Metastatic Adenocarcinoma | MA-11 | Bone metastasis | bisphosphonate efficacy; tumor-osteoclast crosstalk in metastases | [45] |

***Histological subtype***

Tumor morphology, its growth pattern, degree of differentiation, and resemblance to normal terminal duct-lobular units (TDLUs), determine if a lesion is *in situ* or invasive, with invasive tumors carrying a higher risk of metastasis.[46] Histological classifications also guide molecular profiling and subsequent targeted therapy selection. This section details the histological diversity of breast cancer and emphasizes clinic-pathological characteristics and correlations. Adenocarcinomas, comprising over 95% of breast cancer, arise from the glandular epithelium of ducts or lobules (**Figure 1**). They are characterized by glandular differentiation and mucin production, the latter being intracellular, as in signet-ring cells, or extracellular, as seen in mucinous carcinomas.[47] Such tumors are subclassified by their site of origin and invasiveness. Ductal carcinoma is the most prevalent breast cancer type, originating in the mammary milk ducts.[48] Ductal and lobular carcinoma *in situ* represent the two main forms of pre-invasive breast adenocarcinomas. Ductal carcinoma in situ (DCIS) originates in the mammary ducts and remains confined to the ductal system, exhibiting a range of architectural patterns such as solid, cribriform, papillary, and micropapillary.[7] Lobular carcinoma in situ (LCIS), by contrast, arises in the terminal ductal lobular units and is characterized by the proliferation of neoplastic cells that distend and fill the acini. LCIS is considered a non-obligate precursor of invasive lobular carcinoma and is classified into three main histological subtypes: classic, pleomorphic, and florid. These variants differ in cytologic features, architectural patterns, and potential biological behavior.

Invasive lobular carcinomas (ILCs) account for 10-15% of breast cancer cases, arise from the terminal duct-lobular units, and are marked by the loss of E-cadherin, typically due to CDH1 mutations.[49] They are commonly harder to diagnose from mammograms due to not forming calcifications. ILC is usually ER-positive and HER2-negative, with distinct genomic alterations.[50] The majority of ILCs consist of low-nuclear-grade malignant cells [50] (i.e., classic ILCs). In a minority of ILCs, the tumor consists of high-nuclear-grade malignant cells (i.e., pleomorphic ILC). ILC can exhibit a range of histological growth patterns, including solid, alveolar, trabecular, and tubulolobular variants. [49] These patterns reflect the morphological diversity of ILC and occasionally pose diagnostic challenges. Regardless of pattern, these tumors typically retain the hallmark feature of E-cadherin loss, confirming their lobular origin.[29,50] Invasive Ductal Carcinoma (IDC), which constitutes 70–80% of invasive breast cancer cases, invades the stroma and causes desmoplastic reactions.[48] IDC is molecularly heterogeneous, with luminal subtypes expressing hormone receptors, HER2-enriched tumors exhibiting ERBB2 amplification, and basal-like tumors being triple-negative.[7,37] The majority IDCs are classified as no special type (IDC-NST), yet several distinct histological and clinical variants exist.[48]

Metaplastic carcinoma (MpC) is an aggressive form of invasive breast cancer, often classified as triple-negative.[7] It is highly heterogeneous, typified by epithelial-to-mesenchymal transition, which produces variable differentiation including squamous, spindle, or chondroid elements.[51] This subtype of breast cancer is noted for its resistance to chemotherapy.[52] Tubular carcinoma (TC) is a rare form of breast cancer defined by the proliferation of angulated, oval or elongated tubules reminiscent of normal breast ducts. Its invasive nature, coupled with the absence of myoepithelial cells, distinguishes it from benign lesions.[53] Micropapillary carcinoma (MiC) is another aggressive subtype seen in 1–2% of breast cancer cases, characterized by clusters of tumor cells arranged in an inside-out pattern without fibrovascular cores.[54] Despite often being ER-positive, this cancer type display high rates of lymph node involvement and commonly exhibits HER2 amplification or PIK3CA mutations.[54] Adenoid cystic carcinoma (ACC) is an extremely rare subtype (<0.1% incidence), featuring biphasic cell populations, luminal and basaloid, that form tubular, cribriform, or solid patterns surrounded by mucinous material.[7,55] As opposed to its salivary gland counterpart, breast ACC rarely metastasizes, with surgical excision often proving curative.[56]

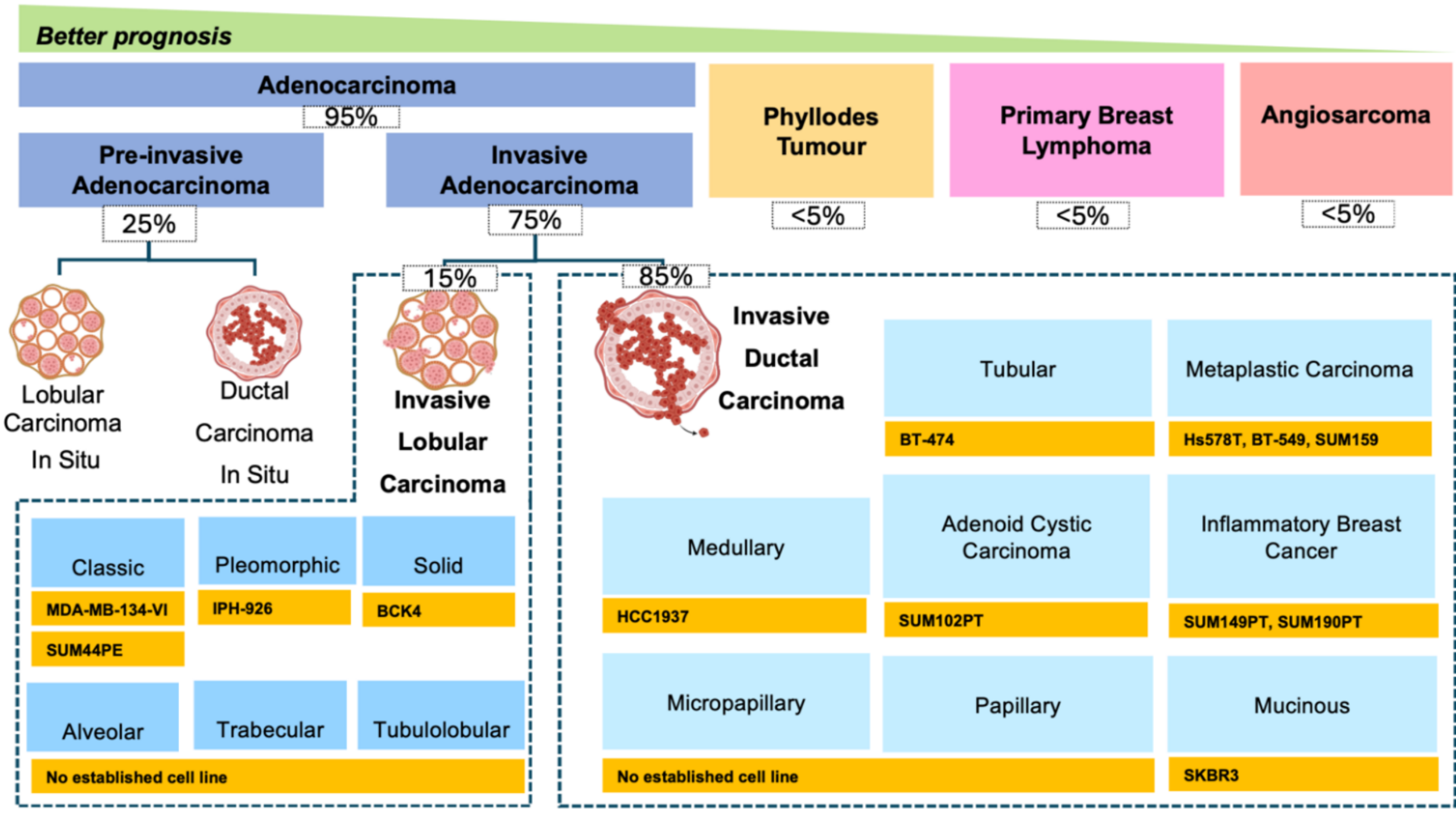

**Figure 1. Histological subtypes of breast cancer.** Breast cancer cell lines are grouped by the histological subtype of the deriving tumors and their sub-classifications, alongside examples of widely used cell lines for each cancer type. For example, SUM149PT and SUM190PT were established from inflammatory breast cancer, while MDA-MB-134-VI and SUM44PE originated from classic ILC. However, not all subtypes are well-represented by directly derived models. In such cases, some widely employed cell lines (e.g., MCF-7 and MDA-MB-231), though not derived from rare subtypes like papillary or metaplastic carcinoma, may still serve as functional models due to their phenotypic behavior.

### *Hormone-receptor status*

Receptors are proteins typically found in the cell membrane that can be bound by matching extracellular molecules to elicit intracellular signaling or to enable inter-cellular communication.[57,58] Some breast cancer cells possess certain receptors to hormones (HRs) that contribute to cellular behaviors including growth, proliferation, and motility.[59,60] HR status has been widely used to classify breast cancer cell lines (**Figure 2**). HRs include the estrogen receptor (ER) and the progesterone receptor (PR). Another important receptor is the human epidermal growth factor receptor 2 (HER2). HR/HER2 expression, among other variables, is one of the most important factors in estimating the prognosis and therapeutic responses of breast cancer.[61] Estrogen receptor-positive (ER+) breast cancer is the most frequently diagnosed subtype. However, only about 30% of the commercially available breast cancer cell lines are ER+, and these models frequently derive from advanced disease states.[62] From those, very few can be grown in mice, such as MCF7 and T47D and ZR-75-1, requiring high levels of exogenous estrogen (E2).[8] This does not reflect the low levels of estrogen found in postmenopausal women, where most cases of ER+ breast cancer develop, making these models limited in scope. The majority of ER+ breast cancer is also PR-positive (PR+).[57] Elevated PR levels are predominantly observed in luminal A tumors, which yield better outcomes compared to luminal B tumors where PR expression is lower.[63] Approximately 15% of breast cancers are human epidermal growth factor receptor 2 positive (HER2+), a subtype that typically affects younger patients and is diagnosed at advanced stages.[64] HER2 overexpression, an independent predictor of poor survival, often occurs irrespective of ER and PR expression.[16] Triple-positive breast cancer (TPBC) is a luminal B subtype co-expressing ER, PR, and HER2, accounting for roughly 10–15% of cases.[65] It often demonstrates suboptimal responses to standard chemotherapy and hormone therapy due to intricate crosstalk between the ER and HER2 pathways.[66] Triple-Negative Breast Cancer (TNBC) lacks expression of ER, PR, and HER2, and accounts for approximately 15% of cases.[67] TNBC is predominantly basal-like, is more common in younger women, and exhibits an increased risk of early recurrence and distant metastasis. It is strongly associated with BRCA1 mutations.[68] The ER+/HER2– subtype represents the most common breast cancer phenotype (approximately 75% of cases) and is typically classified as luminal A-like,[69] while ER+/PR+/HER2+ pattern is classified as luminal B-like.[14] Luminal B cancers with an ER+/PR–/HER2+ profile generally portend a worse prognosis than their ER+/PR+ counterparts, while HER2-positive cancers that are ER–/PR– are managed predominantly with HER2-targeted.[70]

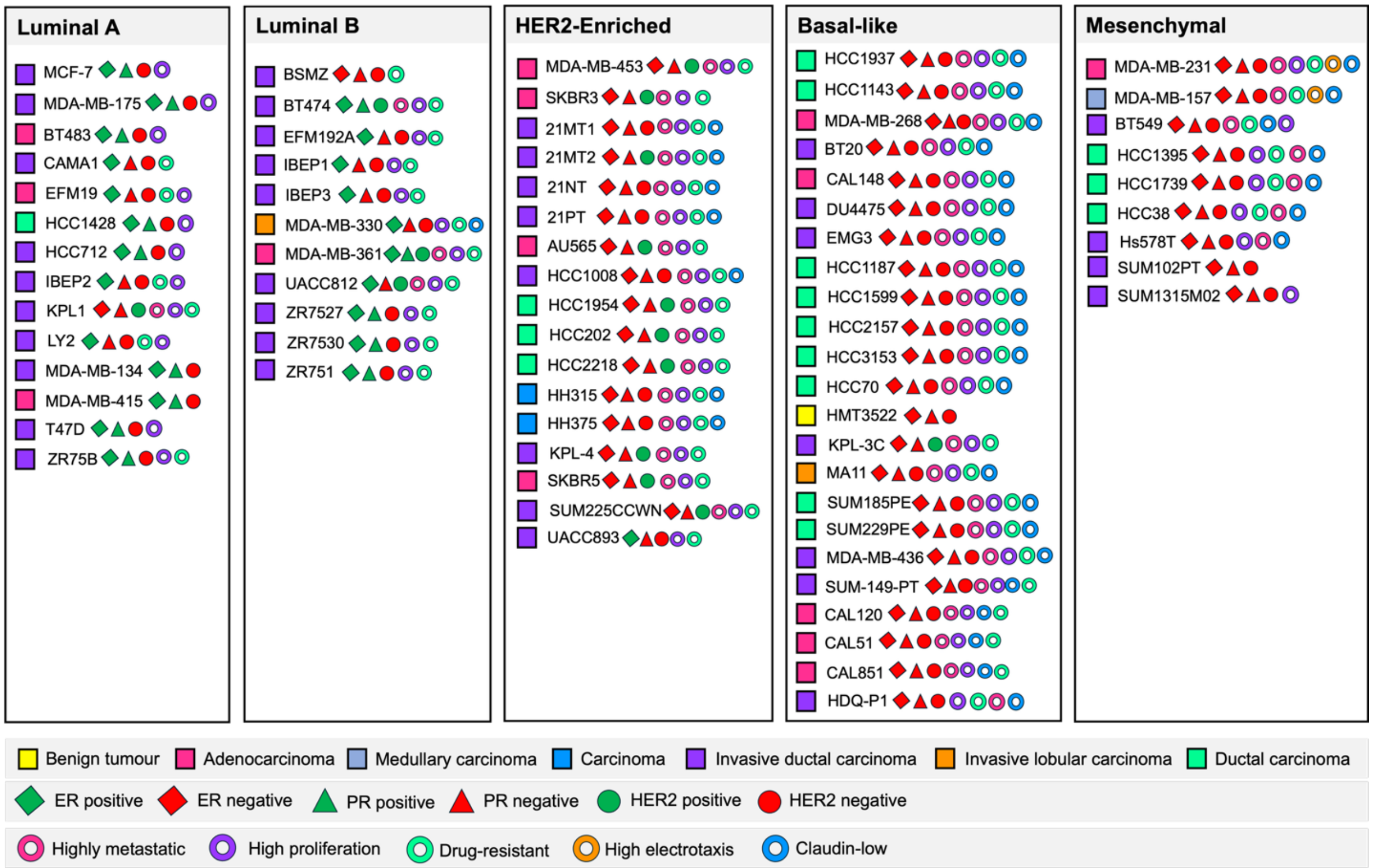

**Figure 2. Classification of selected commercially available human breast cancer cell lines.** The schematic organizes cell lines into five intrinsic subtypes (Luminal A, Luminal B, HER2-enriched, Basal-like, and Mesenchymal) displayed in columns. Color-coded symbols represent key attributes used in absolute and relative classifications previously discussed hereby to map their molecular and functional characteristics.

### *Genetic mutations*

Breast cancer is primarily driven by genetic factors, with age and family history being the most significant risk factors. Approximately 5-10% of breast cancer cases are associated with inherited gene mutations.[10,71] Germline alterations in BRCA1 and BRCA2, among the most widely known mutations, compromise DNA repair and confer a markedly increased lifetime risk, predisposing tumors to either triple-negative or predominantly ER-positive phenotypes, respectively.[72] Mutations in TP53, present in nearly 30% of cases, disrupt critical cell cycle checkpoints and promote aggressive tumor behavior with poorer outcomes.[9] PTEN mutations are strongly correlated with HER2+ breast cancers,[73] and inversely associated with luminal type breast cancers[74] affecting cell growth, and proliferation, and inhibiting cancer stem cell activity.[75] Other important mutations include defects in CHEK2 and ATM weakening cell cycle control and apoptotic responses.[76] Alterations in PALB2, CDH1, STK11, and NF1 contribute to genomic instability, drive invasive characteristics, and influence therapeutic resistance[77]. Genetic aberrations define distinct molecular subtypes in breast cancer and are essential for guiding targeted treatments in precision oncology. Genetic drift represents a different problem, further addressed in this review.

### *Molecular subtype*

Gene expression profiling and hierarchical clustering have delineated five principal molecular subtypes, each with distinct biological behavior, risk factors, and therapeutic responsiveness, namely luminal A, luminal B, HER2-enriched, basal-like, and claudin-low.[78] Luminal A is the most common subtype of breast cancer, accounting for around 40% of all breast cancer cases.[78] It is characterized by an expression of luminal (low molecular weight) cytokeratins, ER, and PR, with a HER2 negative profile accounting for the low expression of cell proliferation marker Ki-67 (less than 20%).[79] Luminal B subtype represents 20–30% of cases, expressing ER (with often reduced PR) and displaying high proliferation indices (Ki67 above 20%).[79] These are generally of higher histologic grade, more aggressive, and have a higher recurrence rate compared to Luminal A subtypes, necessitating combined endocrine and chemotherapeutic approaches.[84] HER2-enriched comprise approximately 15% of breast cancer cases.[80] These are HER2-positive tumors while often exhibiting low or absent ER and PR levels.[81] This category is subdivided into luminal HER2 (E+, PR+, HER2+ with intermediate Ki67, 15–30%) and HER2-enriched (E–, PR–, HER2+ with high Ki67, >30%), both marked by high-grade invasive ductal carcinomas with nodal positivity and aggressive clinical behavior.[78] Basal-like is often used as a synonymous of triple-negative breast cancer (TNBC), lacking ER, PR, and HER2 expression, while expressing basal cytokeratins.[82] Basal-like breast cancers are typically high grade, occurring in patients with BRCA1 mutations, and have limited treatment options outside chemotherapy. Claudin-low tumors are characterized by low expression of cell adhesion molecules and a stem cell–like phenotype. This rare and aggressive subtype is often considered a subclass of basal-like but has gained interest as an *in vitro* model to reproduce highly aggressive cancers.[52]

#### *Patient age, gender, and ethnicity*

Additional absolute criteria include patient age, gender, and ethnicity. Age critically influences breast cancer risk, tumor morphology, and treatment response.[78] Tumors in patients under 40 typically exhibit reduced levels of estrogen receptor, progesterone receptor, and luminal cytokeratin, alongside elevated Ki67, HER2, and p53 expression, indicative of aggressive behavior.[83] In contrast, tumors in individuals over 70 generally display indolent features. Yet, most commercially available cell lines were derived from older patients, potentially limiting experimental relevance.[83] Ethnicity further modulates breast cancer biology, as disparities in healthcare result in later diagnoses in Hispanic and Asian populations, while non-Hispanic black patients exhibit a tumor microenvironment enriched with pro-tumorigenic immune cells, enhanced microvasculature, and elevated mitotic kinases and transcription factors that promote aneuploidy.[86] More in general, marketed cell lines are predominantly caucasian.[84] With regards to gender, although breast cancer is most commonly viewed as a female disease, it can also occur in men, where it accounts for less than 1% of all cancers in men and breast cancer cases overall.[85] However, male breast cancer incidence has rose over the past 30 years, with inherited pathogenic variants being the most significant risk factors.[86] Transgender individuals may also face breast cancer risks, particularly if receiving hormone treatment. Studies have shown an increased risk of breast cancer in transexual women compared with cisgender men, and a lower risk in trans men compared with cisgender women.[87]

### RELATIVE CLASSIFICATION CRITERIA

This section details key classification parameters specific to breast cancer to emphasize their mechanistic underpinnings and clinical utility. Such features are defined as relative due to providing a qualitative measure of cancer cell behavior. Important relative criteria include metastatic ability, proliferation rate, often measured by Ki-67 expression or mitotic indices,[88] response to radiation reflecting the tumor sensitivity to DNA damage-induced cell death,[89] and drug resistance encompassing mechanisms by which tumors evade therapeutic agents (such as ESR1 mutations in hormone-resistant HR+ disease).[90] Inflammatory status reflects, among all, immune microenvironment composition and stem cell-like properties.[91] A summary of key relative criteria, their working mechanisms and implications are summarized in **Table 2**, alongside absolute criteria to provide a complete overview.

#### *Metastatic ability*

Metastatic ability refers to the capacity of tumor cells to colonize distant organs such as bone, brain, and liver, and it is driven by EMT.[11] In metastatic behavior, transcription factors downregulate E-cadherin, enhancing motility and favoring migration to colonize new sites. Bone metastasis, one of the hardest to treat, involves osteolytic factors that activate osteoclasts via RANKL signaling.[11] Circulating tumor cells expressing HER2 or EpCAM have been associated with increased metastatic risk.[92] Metastatic behavior also differs based on subtype-specific organotropism: luminal tumors often metastasize to bone, HER2+ to liver and lungs, and basal-like tumors to brain and lung[12]. These preferences reflect intrinsic properties of the tumor cells, including receptor expression, secreted factors, and ability to modify the pre-metastatic niche. Some cell lines, like MDA-MB-231, consistently metastasize to lung and bone in murine models, while others such as MCF-7 require estrogen supplementation and genetic manipulation to become metastatic.[93] Additional drivers include specific signaling cascades, for example CXCR4-CXCL12 involved in bone homing, and MMPs, which degrade extracellular matrix to facilitate invasion.[11] Exosomes from metastatic cells can precondition distant niches to favor colonization.[12] Moreover, stem-like subpopulations with $CD44^{high}/CD24^{low}$ profiles exhibit heightened metastatic ability and are linked to relapse.[19,52]

### *Migration and invasiveness*

Migration and invasiveness describe the ability of cancer cells to detach from the primary tumor site, degrade the extracellular matrix, and infiltrate surrounding tissues. [12] Invasive breast cancer, particularly TNBC and HER2-positive subtypes, exhibit heightened migration via RhoA/ROCK-mediated cytoskeletal reorganization and MMP-9/MMP-14-dependent extracellular matrix degradation.[94] *In vitro* models, for instance MDA-MB-231 cell invasion assays, reveal that TGF-β signaling enhances motility by upregulating integrin αvβ6. [82] These behaviors are often initiated during partial EMT which increases cellular plasticity while maintaining some epithelial traits, allowing dynamic adaptation to microenvironmental cues. Migratory activity can be random or directional (chemotaxis), with directional migration often guided by gradients of stromal-derived factors.[95] Tumor-associated fibroblasts and macrophages also promote invasion by remodeling the ECM and secreting pro-migratory cytokines[96]. Additionally, invadopodia (actin-rich protrusions) formation facilitates local matrix degradation and is prominent in highly invasive cell lines. [69]

### *Apoptotic resistance*

Apoptotic resistance denotes the tumor evasion of programmed cell death, enabling survival despite genomic damage or therapy.[13,90,97]Apoptotic evasion in breast cancer is linked to Bcl-2 overexpression in luminal subtypes and TP53 mutations in basal-like tumors.[35,82] TNBCs frequently exhibit FLIP upregulation, which inhibits caspase-8 activation.[95] PARP inhibitors (e.g., olaparib) exploit synthetic lethality in BRCA1/2-mutated tumors by impairing DNA repair and forcing apoptosis.[98] Some breast cancer cells bypass apoptosis entirely by entering a senescent-like state or activating autophagy as an adaptive survival strategy under therapeutic stress[15]. Apoptotic resistance is commonly assessed *in vitro* using Annexin V/PI staining, TUNEL assays, and caspase-3/7 activity measurements.[5,13] Cell lines like MDA-MB-231 and BT-549 are often used to model high apoptotic resistance, particularly in response to chemotherapy or radiation[15].

### *Gene expression profile*

Gene expression profiles represent the transcriptomic signatures that classify breast cancer into intrinsic subtypes. Intrinsic subtypes luminal A, luminal B, HER2-enriched, and basal-like, originated from breast cancer transcriptomics.[99] The PAM50 assay further refines classification, identifying a claudin-low subgroup with stem-like features.[82] Oncotype DX and MammaPrint quantify recurrence risk using proliferation (e.g., Ki-67) and invasion-related genes (e.g., MMP11).[100] As a relative criterion, gene expression profiling enables comparison of cell lines beyond subtype labels, revealing functional differences in pathway activity, hormone signaling, immune evasion, or stemness. For instance, two basal-like cell lines may diverge significantly in EMT gene signatures or interferon response genes, making one of them more suitable for metastasis or immunotherapy studies. Expression levels of DNA repair genes (e.g., BRCA1, RAD51), growth factors (e.g., TGFB1), or apoptotic regulators (e.g., BCL2, CASP9) offer insights into therapeutic vulnerabilities[101]. Additionally, cell lines cultured in 2D versus 3D conditions can exhibit shifts in gene expression[102], highlighting the need to interpret transcriptomic data in context and developing *in vivo*-like models to improve accuracy.

### *Epigenomic modifications*

Epigenomic modifications, for example DNA hypermethylation of BRCA1 occurring in a significant fraction of sporadic tumors, are reversible with HDAC inhibitors that restore ERα expression, while EZH2 overexpression in TNBC correlates with stemness and metastasis.[103] In a model selection context, epigenomic features help distinguish cell lines by their regulatory landscape rather than transcriptional output alone. DNA methylation at tumor suppressor *loci* or enhancer regions can create long-term silencing that persists across treatments, as well as histone modifications that can influence chromatin accessibility and lineage identity.[97] Functional assays using 3D organoids and patient-derived xenografts provide dynamic insights into treatment response and resistance.[97] Cell lines differ in the stability of these states, some retaining locked epigenomes, while others are more prone to reprogramming, particularly in 3D culture or co-culture with stromal cells.[102]

### *Tumor-immune interactions*

Tumor-immune interactions are one of the most important features of cancer cells when fighting against therapies and initiating invasion. Cancer cells can evade immune cells and recruit them as tumor-associated macrophages, regulatory T cells or myeloid-derived suppressor cells, further exacerbating the malignancy of the tumor[96]. Even typically anti-tumor cells (e.g., T cells or NK cells) can become exhausted or suppressed in the TME due to interaction with specific factors, hypoxia, or the absence of stimulatory signals.[96] Some breast cancer cell lines, for example MDA-MB-231, HCC38, HCC70 and BT-549, can recapitulate this interaction, namely, cells with high immunogenicity.[104] These are often highly modulated when co-cultured with macrophages or T cells and express high levels of cytokines, chemokines, and high PD-L1 under IFN-γ stimulation.[96] Low immunogenic cell lines, such as MCF-7, T47D and ZR-75-1, lack expression of PD-L1 and have a low response to IFN- γ or immune co-cultures, making them less suitable for immune checkpoint or immunotherapies studies.[16] However, even high-interaction lines are still imperfect models, as standard cell lines lack the full complexity observed in immune microenvironments *in vivo*.[105]

**Table 2. Absolute and relative classification criteria, key features, and clinical relevance.**

| | | Breast cancer subcategory | Key features & molecular details | Clinical relevance | Ref. |
|---|---|---|---|---|---|
| Absolute criteria | Histological subtype | *Adenocarcinoma* | >95% of cases; arises from glandular epithelium; glandular differentiation and mucin production | Subtyped as ductal vs. lobular; informs targeted therapy | 24,47 |
| | | *Ductal carcinoma* | Divided into DCIS and IDC; IDC is molecularly heterogeneous (luminal, HER2, basal-like) | Provides prognostic stratification based on grade and subtype | 7,48 |
| | | *Lobular carcinoma* | Arises from TDLUs; loss of E-cadherin (CDH1 mutations); usually ER+ and HER2–; associated with FOXA1, TBX3 mutations | Complicates detection; influences therapeutic strategies | 49,50 |
| | | *Inflammatory breast cancer* | Rare (1–5%); typically, triple-negative or HER2+; overexpresses EGFR, ANXA1, and COX-2; activation of WNT/β-catenin and NF-κB pathways | Highly aggressive with rapid progression | 106 |
| | | *Medullary carcinoma* | Syncytial growth (>75%), absence of glandular/tubular structures; frequent mitoses | Rare IDC variant with distinct histological features | 7 |
| | | *Mucinous carcinoma* | Extracellular mucin; clusters; typically, ER+, HER2–; low TP53 mutation; AKT1 E17K mutations | Generally lower grade and favorable prognosis | 107 |

| | | | | |
|---|---|---|---|---|
| | *Papillary carcinoma* | Papillary with fibrovascular cores; subtypes include intraductal, encapsulated, solid, and invasive forms | Crucial to differentiate benign from malignant lesions | 108 |
| | *Metaplastic carcinoma* | Aggressive TNBC subtype; heterogeneous with evidence of EMT transition; chemoresistant | High resistance profiles; therapeutic challenges | 51,52 |
| | *Tubular carcinoma* | Well-differentiated; small cell tubules arranged radially; invasive | Rare, low-grade, and excellent prognosis | 53 |
| | *Micropapillary carcinoma* | Often ER+ with high lymph node metastasis; MUC1 overexpression; may have HER2 amplification or PIK3CA mutations | Poorer prognosis necessitating adjuvant chemotherapy | 109 |
| | *Adenoid cystic carcinoma* | Rare (<0.1%); TN yet indolent; MYB-NFIB fusions triggering NOTCH pathway activation | Surgical excision is often curative | 56 |
| **Hormone Receptor Status** | *ER+* | Most prevalent; includes MCF7, T47D, ZR-75-1; require high exogenous estrogen; responsive to endocrine therapy | Cell lines may not mimic low estrogen conditions of postmenopausal patients | 57 |
| | *PR+* | Expressed in response to ER activation; higher levels common in luminal A; prognostic marker | PR positivity generally correlates with better outcomes | 58 |
| | *HER2+* | 15% of cases; overexpression of HER2; adverse prognostic indicator independent of ER/PR | Managed with HER2-targeted agents (e.g., trastuzumab) | 64 |
| | *Triple positive (TPBC)* | Co-expression of ER, PR, and HER2; luminal B subtype (~10–15%); pathway crosstalk | Requires combinatorial therapeutic approaches | 66 |
| | *Triple negative (TNBC)* | Lacks ER, PR, and HER2; predominantly basal-like; distant metastasis; linked with BRCA1 mutations | Limited targeted therapies | 59 |
| | *ER+/HER2–* | Most common phenotype (~75% of cases); classified as luminal A | Endocrine treatments | 69 |

| | | | | | |
|---|---|---|---|---|---|
| | | *ER+/PR–/HER2+* | Luminal B variant with altered receptor signaling; poorer prognosis compared to ER+/PR+HER2+ | Potential endocrine therapy resistance | 70 |
| | | *ER–/PR–/HER2+* | HER2-overexpressing cancers lacking hormone receptors | Treated with HER2-targeted drugs | 80 |
| | Genetic Mutations | *BRCA1* | Germline mutations (16% of hereditary BC); crucial for DNA repair and cell cycle regulation; TNBC | Key target for PARP inhibitor strategies | 110 |
| | | *BRCA2* | Functions in DNA repair and genomic stability; 70–80% of BRCA2-mutated cancers are ER+ | Influences endocrine therapy in mutation carriers | 72 |
| | | *TP53* | Mutated in ~30% of BC; mediates cell cycle arrest, apoptosis, or senescence upon DNA damage | Determines tumor aggressiveness | 9 |
| | | *PTEN* | Regulates the PI3K/Akt pathway; linked with HER2+ cancers, inversely with luminal types | Its loss may direct targeted treatment strategies | 74 |
| | | *PALB2* | Works in tandem with BRCA1/2 in DNA repair; often associated with aggressive TNBC phenotypes | Emerging target in personalized therapeutic approaches | 111 |
| | | *CHEK2* | Checkpoint kinase mutation; majority are luminal A with some lobular features | Reflects cell cycle checkpoint dysfunction | 112 |
| | | *ATM* | Involved in cell cycle regulation and apoptosis; loss-of-heterozygosity in ~40% of sporadic BC; mainly in luminal B/HER2- | May influence chemotherapeutic decisions. | 113 |
| | | *CDH1* | Encodes E-cadherin; loss leads to cell dissociation; mutations predispose to invasive lobular carcinoma | Essential for hereditary lobular BC surveillance | 114 |
| | | *STK11* | Associated with Peutz-Jeghers syndrome; predisposes to other tumor types (ovarian, lung, GI) | Important for multi-tumor screening | 115 |
| | | *NF1* | Mutations in NF1 occur in ~27% of BC; endocrine resistance and metastasis, especially in ER+ cases | May predict targeted endocrine resistance management | 81 |
| | **Molecular Subtype** | *Luminal A* | ~40% of cases; ER/PR positive; low Ki67 (<20%); typically, IDC; high endocrine therapy response | Represents the best prognostic subgroup | 79 |
| | **Molecular Subtype** | *Luminal B* | 20–30% of cases; ER positive (with lower PR expression); high Ki67 (>20%); variable HER2 | More aggressive; requires combined therapy approaches | 107 |
| | **Molecular Subtype** | *HER2-enriched* | ~15% of cases; overexpresses HER2; subdivided into luminal HER2 and HER2-enriched | Aggressive; outcomes improve with HER2-targeted therapies | 78 |
| | **Molecular Subtype** | *Basal-like (TNBC)* | 15–20% of cases; lacks ER, PR, and HER2; expresses basal cytokeratins; linked with BRCA1 | Limited treatment options | 78 |
| | **Molecular Subtype** | *Mesenchymal (claudin-low)* | Rare; low expression of cell adhesion molecules and stem-cell-like phenotype | High tumor plasticity and potential resistance | 81 |
| **Relative Criteria** | **Metastatic ability** | | EMT transition; downregulation of adhesion molecules; organ tropism. SNAIL, TWIST, ZEB1; downregulation of E-cadherin; PTHrP, IL-11; RANKL; HER2, EpCAM; overexpression of COX2, HB-EGF, ST6GALNAC5 in brain-tropic cancers | Enhanced motility; colonizes distant sites | 11,92 |
| **Relative Criteria** | **Tumor, migration and invasiveness** | | Cell detachment; ECM degradation; cytoskeletal reorganization; invadopodia formation; integrin-mediated motility. MMP-9, MMP-14; RhoA/ROCK; TGF-β–driven integrin αvβ6; proliferation markers like Ki-67; Cyclin D1–CDK4/6 complexes; MYC amplification; PI3K/AKT hyperactivity | Invasive capacity; differential proliferation in luminal A vs. B subtypes; targeted therapeutic strategies | 88,94 |

| | | | | |
|---|---|---|---|---|
| | Apoptotic resistance | Upregulation of anti-apoptotic proteins; Overexpression of Bcl-2 (luminal tumors), TP53 mutations (basal-like), FLIP upregulation in TNBC; BRCA1/2; HIF-1α and carbonic anhydrase IX in hypoxic conditions | Chemo/radio resistance; combined treatments such as PARP inhibition with radiotherapy | |
| | Gene expression and drug resistance | Resistance through specific mutations. PAM50, Oncotype DX, MammaPrint; ESR1 mutations (Y537S, D538G); PTEN loss; PIK3CA mutations; βIII-tubulin overexpression; sensitivity to HER2-targeted agents, platinum salts, and AR inhibitors | Recurrence risk assessment; endocrine and targeted therapy resistance | 116 |
| | Epigenomic modifications | BRCA1 hypermethylation reversible by HDAC inhibitors; EZH2 | Influences chemoresistance, recurrence, and therapeutic response | 30 |
| | Tumour - immune system interactions | Tumour-infiltrating lymphocytes; IL-6/STAT3-mediated PD-L1 upregulation; ALDH1, CD44/CD24; Notch, Hedgehog, WNT/β-catenin; HR deficiency scores; APOBEC3B | Immunotherapeutic resistance and increased invasive capacity | 16,105 |

## TOWARD INFORMED CELL MODEL SELECTION

To consolidate the criteria discussed throughout this review and translate them into actionable steps, this section proposes a structured workflow for breast cancer cell line selection (**Figure 3**). Starting from a defined research question and primary goal, whether mechanistic exploration, drug development, biomarker discovery, or modeling tumor–microenvironment interactions, the approach can guide users through a series of filters based on absolute and relative criteria. The layers include fundamental features followed by dynamic functional attributes relevant to experimental aims. Across levels, the candidate pool progressively narrows down to lines that are, however possible, biologically appropriate, experimentally feasible in the given context, and clinically relevant to their conceived application. The workflow further integrates downstream considerations, for example validation needs (e.g., passage number, genetic drift), with suggestions for additional experimental steps before proceeding, as well as representation gaps (e.g., age range of applicability). Crucially, it also introduces the concept of scaling potential, i.e. the likelihood that a given cell line will perform well in advanced disease models such as organoids, co-culture systems, organ-on-chip platforms or patient-derived xenografts. Pre-selecting based on functional and phenotypic compatibility to more advanced platforms can reduce resource misuse and improve translational fidelity, possibly aiding toward avoiding the use of unscalable *in vitro* models in studies originally conceived for *in vivo* applications. This structure might lay the foundation for building adaptive and AI-supported tools that link cell line metadata with model performance to optimize preclinical design.

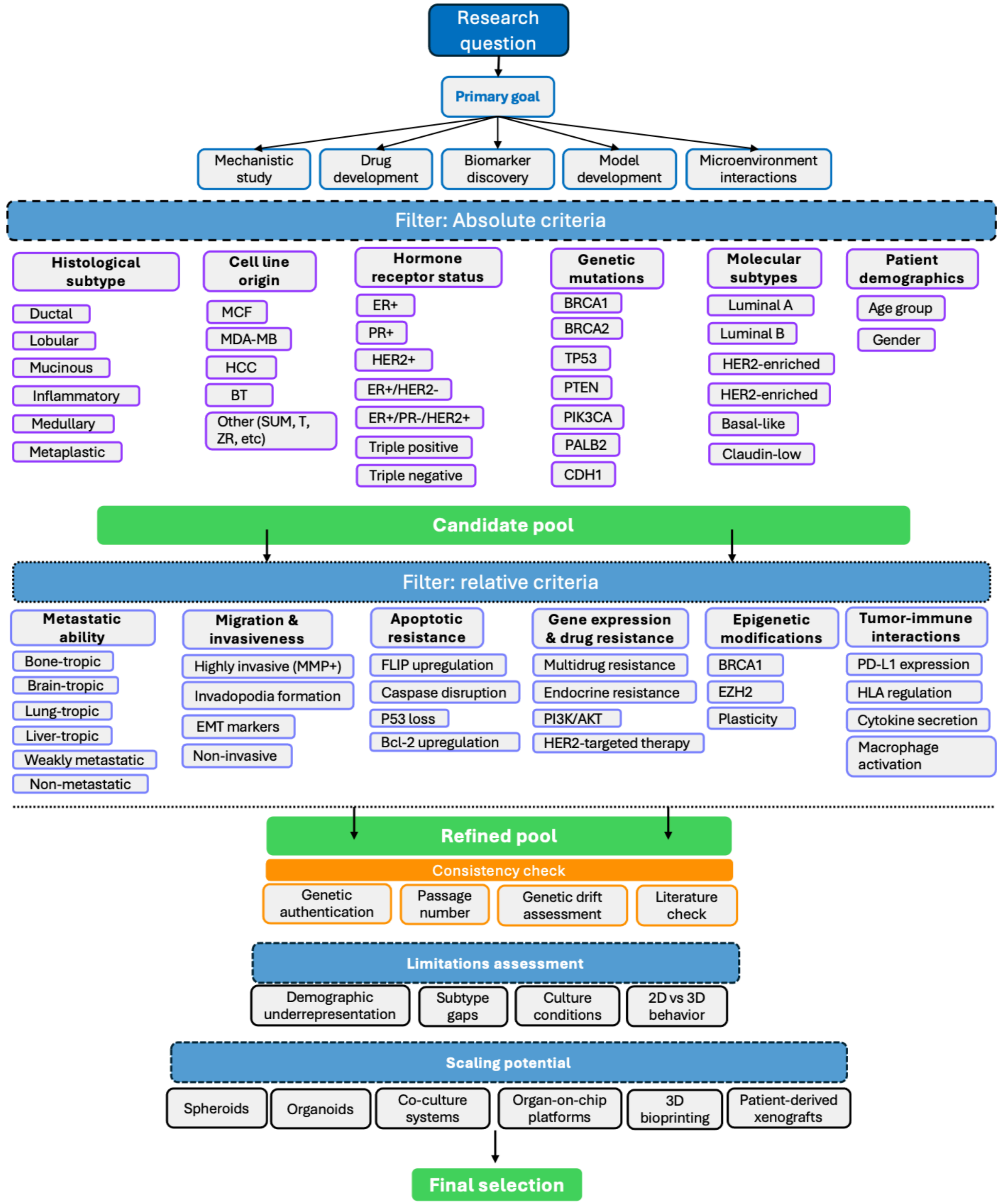

**Figure 3. Decision-making workflow for breast cancer cell line selection.** The diagram outlines the proposed stepwise filtering process, beginning with the research objective. It then guides selection through sequential filtering criteria: absolute filters (e.g., receptor status, molecular subtype), and relative filters (e.g., metastatic tropism, resistance traits). Additional layers incorporate experimental feasibility, validation requirements (e.g., passage history, genetic drift), and representation gaps. The final step considers the scalability of selected lines toward advanced disease models such as organoids, co-cultures, and xenografts.

The practical value of this approach is presented in two research scenarios requiring thoughtful cell line selection (**Figure 4**). In the first case, investigating bone-specific metastasis mechanisms in ER+ breast cancer through a standard literature search or AI chat assistant might yield conflicting results, with studies using MCF-7 cells despite their limited metastatic capacity or MDA-MB-231 cells despite being ER-negative.[117] By applying the presented model, one would first filter by absolute criteria (ER+ status) and then by relative criteria (bone-metastatic potential), leading to the selection of MCF-7-derived bone-seeking variants or ZR-75-1 cells, which express PTHrP and other bone-metastasis mediators while maintaining ER positivity.[90]

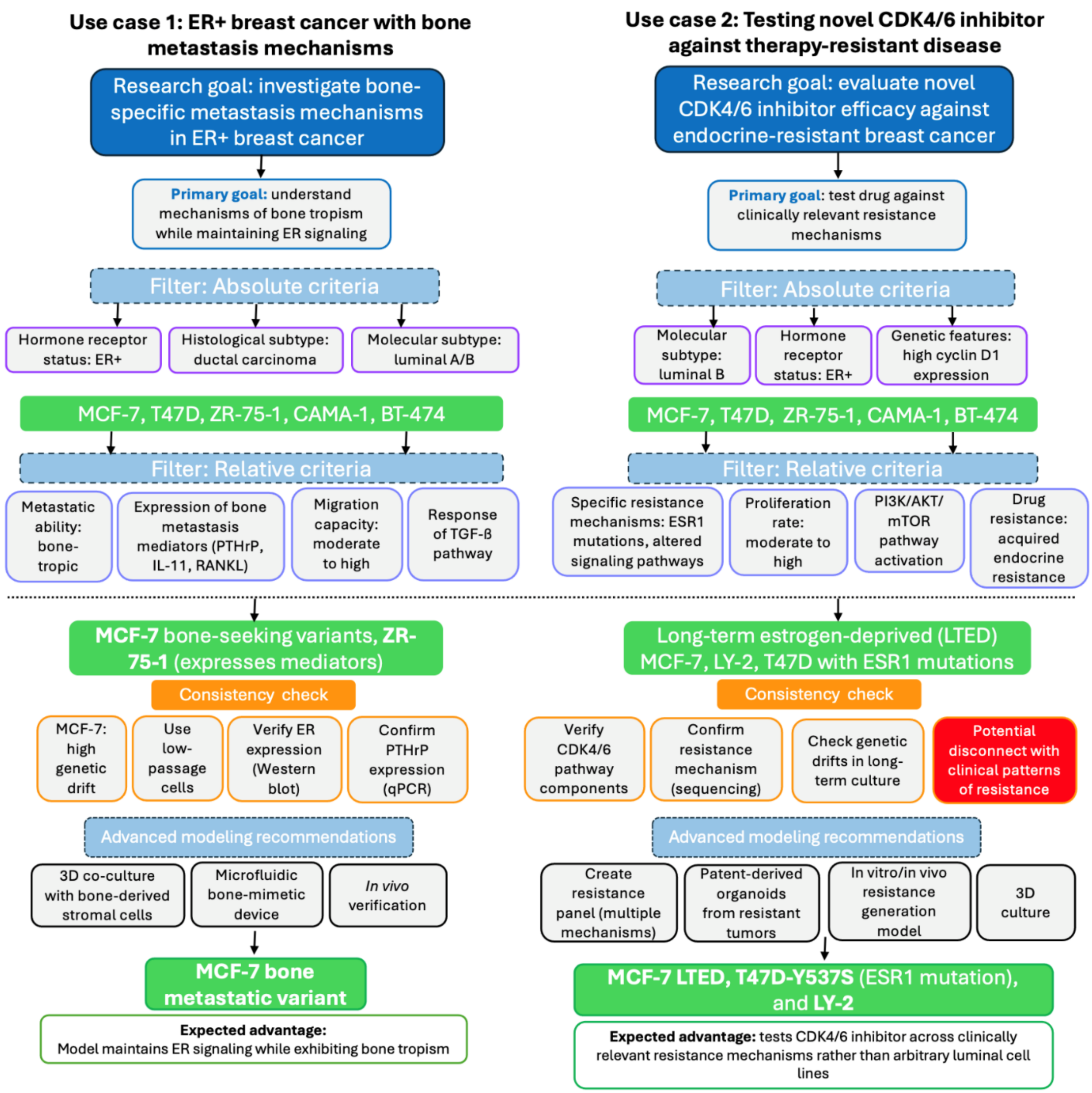

**Figure 4. Use cases of the proposed workflow.** The first use case (left panel) addresses bone metastasis in ER+ disease, identifying optimal cell models by filtering for both receptor status and metastatic competence. The second (right panel) illustrates drug testing for therapy resistance, combining genomic features with acquired phenotypes to select resistant luminal B models. In both cases, this approach could outperform standard selection methods by working with contextual fidelity. It also integrates considerations of genetic drift and validation needs across strains to potentially add experimental validation steps or pivot.

The second sample use case presented here involves a novel CDK4/6 inhibitor being tested against therapy-resistant disease. Instead of arbitrarily selecting a panel of luminal cell lines, the framework guides the combination of absolute criteria (luminal B classification, high Cyclin D1 expression) with relative criteria (acquired endocrine resistance). This would lead to selection of long-term estrogen-deprived (LTED) MCF-7 derivatives, LY-2 cells with acquired tamoxifen resistance, or T47D cells with ESR1 mutations, providing a physiologically relevant resistance context.[90] The consideration of genetic drift would alert potential inconsistencies between different laboratory strains and encourage authentication and early-passage usage. In both cases, this approach might deliver contextually relevant models that mirror specific disease states over generic categorizations, to enable translational outcomes compared with a simple database consultation or literature search.

## INTEGRATION WITH ADVANCED AND EMERGING DISEASE MODELS

The development and deployment of disease models aiming to recapitulate breast cancer complexity benefit from the integration of the defined criteria into the set of advanced and emerging technologies, for example multi-omics profiling to enhance model fidelity, machine learning to forecast treatment outcomes, and microsystems mimicking tissue microenvironments (**Figure 5**).[33,36,105,118] Integrated analyses show that many breast cancer cell lines cluster into familiar intrinsic subtypes (e.g. luminal vs. basal) based on gene expression, mirroring patient tumor classes, and multi-omics approaches have even revised prior cell line classifications.[36,47] The combination of genomics (mutations, copy-number variation), transcriptomics (mRNA, miRNA), epigenomics, and proteomics allows us to obtain a holistic molecular portrait that improves subtype matching. This approach also enables to spot discrepancies, as breast cell lines often carry more numerous mutations than tumors, and several key metastatic drivers (e.g. ESR1 mutations) found in patient tumors are absent in commonly used lines.[90,117] Bringing together multi-omics data from patients (e.g. The Cancer Genome Atlas) with cell line and organoid data helps identify which models best recapitulate the molecular wiring of a given subtype, potentially enabling the discovery of hybrid molecular subgroups and biomarkers that are overlooked in single-omics analysis.[67] Multi-omics integration enables us to match absolute features (e.g., BRCA1 mutation, ER status, luminal B signature) with its functional behaviors (e.g., apoptotic resistance, metastatic tropism), as captured by the relative criteria, simulating clinically relevant contexts *in vitro*. For example, a drug targeting CDK4/6 should not solely be tested in a luminal-type cell line, but also in one with high Cyclin D1 expression and low apoptotic priming, integrating a profile that combines static and dynamic features. On top of this, machine learning is increasingly able to predict drug responses and classify breast cancer subtypes with accuracy through algorithms trained to recognize complex patterns that correlate with clinically relevant features. The NCI-DREAM Challenge assembled genomic, transcriptomic, proteomic, DNA methylation, and mutation data for 53 breast cancer cell lines to predict relative sensitivity to various drugs.[119] A recent study built a deep neural network that integrates gene expression, DNA copy number, mutations, and phospho-proteomic (RPPA) data, including a graph-embedded layer of protein–protein interactions.[9] Other groups have developed multi-omics machine learning frameworks to classify breast tumors into subtypes and predict patient therapy outcomes.[120] As a real-world example, ensemble models trained on cell line screens have been applied to patient tumor data to distinguish responder vs. non-responder profiles. [119]

Traditional 2D monolayer cultures often fail to recapitulate the complex architecture and microenvironment of breast tumors, potentially distorting cellular behavior with cells spreading unnaturally, losing their apico-basal polarity, lacking contact with a native ECM, and altering gene expression and drug responses[102,105]. For example, 2D-cultured cells have continuous access to oxygen and nutrients, as opposed to cells in a tumor mass, potentially making them more sensitive to drugs than real tumors. Advanced culture systems like 3D spheroids, patient-derived organoids, co-culture models and 3D bioprinting are now improving the physiological relevance of preclinical studies. 3D cultures allow cells to aggregate or grow in a matrix, restoring more *in vivo*-like morphology and cell–cell interactions. A core benefit of 3D tumor spheroids is the emergence of oxygen, nutrients, and drug gradients across the spheroid, leading to heterogeneous zones of proliferating vs. quiescent cells, mirroring solid tumors.[121] Studies comparing *in vitro* models have shown that breast cancer cells in 3D can become significantly more drug-resistant than in 2D monolayers.[105,118,121] For instance, when MCF-7 spheroids were treated with common chemotherapeutics (doxorubicin, paclitaxel, tamoxifen), the 3D cultures were markedly less responsive than 2D cultures[121]. This *in vitro* drug resistance in 3D is attributed to the restoration of tumor-like cell–cell/ECM interactions and diffusion barriers that limit drug penetration. Organoids are often derived from patient tumor samples or stem cells, that self-organize into mini-structures ideally containing multiple cell types, maintain the DNA copy number aberrations and sequence mutations of the parental tumor, as well as estrogen receptor (ER), HER2, and other subtype-defining features. Organoid biobanks and consortia (e.g., the HUB/Hubrecht Institute and Human Cancer Models Initiative) are now cataloguing large collections of breast cancer organoids for research. [122]

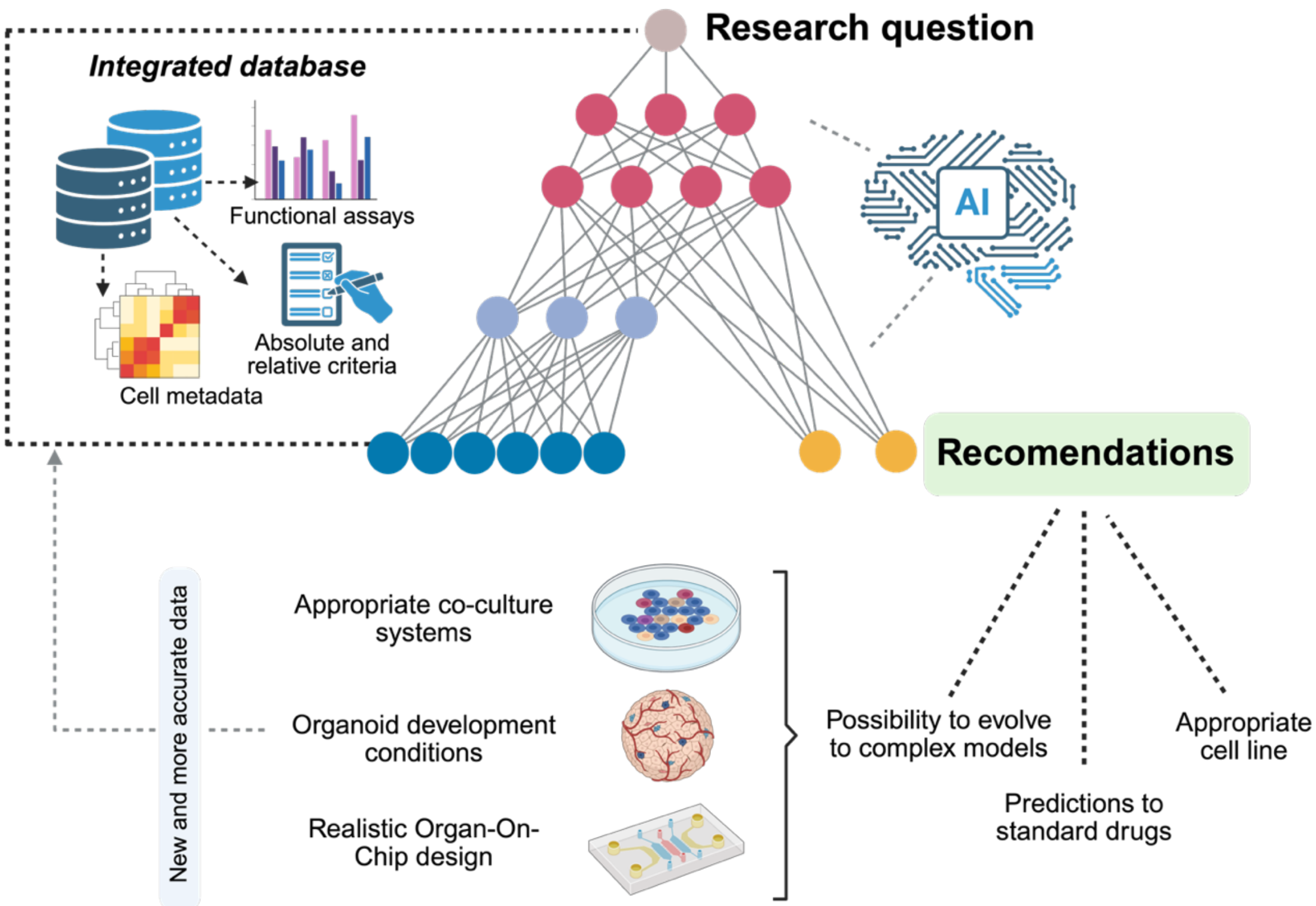

**Figure 5. Sample workflow in AI-assisted optimization of cell line selection for 2D culture and advanced disease models.** Cell line metadata and experimental data are compiled into a structured database, from which data is fed as inputs into a machine learning architecture that integrates absolute and relative criteria to generate predictive recommendations tailored to research goals. Suggested cell lines could be deployed in increasingly complex *in vitro* platforms to enhance biological fidelity and clinical relevance. The resulting experimental outputs, once validated, could feed back into the system, allowing continuous model refinement and improved prediction accuracy over time.

Another approach to breast cancer modelling revolves around the development of co-culture systems, recalling that tumors are ecosystems of cancer cells interacting with stromal cells, immune cells, neurons, and extracellular matrix. Breast cancer spheroids have been co-cultured with cancer-associated fibroblasts or immune cells to reconstitute aspects of the tumor microenvironment.[105] These methods can reveal emergent behaviors, such as how fibroblasts induce drug resistance or how immune cells infiltrate tumor spheroids. A parallel method around the development of advanced systems is organ-on-chip technology, which uses microfabrication techniques to recreate environments that allow blood flow or gradients of nutrients across a biological sample. A study reports the use of a 3D bioprinted breast tumor model in a microfluidic chip, arranging breast cancer cells (MCF7, MDA-MB-231) and healthy mammary cells (MCF10A) within a hydrogel scaffold.[102] This method demonstrated realistic cell migration and invasion patterns in response to gradients. 3D bioprinting, in turn, represents a popular frontier for building custom tumor models that include patient-derived cancer cells, supporting fibroblasts, endothelial cells to mimic blood vessels, and immune components all in one 3D construct. In this context, bringing together cell lines that collectively recapitulate a desired environment becomes more complex to do by hand, as well as crucial for ensuring the validity of experimental outcomes. Then, *in vivo* validation remains a crucial step. A gold standard approach is the use of patient-derived xenografts (PDX), where fragments of a breast tumor are implanted into immunocompromised mice[123]. Studies have shown that if a drug causes regression in a cohort of breast cancer PDX models, there is a good chance it will show efficacy in patients with similar tumor profiles.[118] PDXs are also instrumental in co-clinical trials, i.e. parallel studies where patients receive therapy while their tumor xenografts in mice are treated similarly, acting as a great standard to comprehensively evaluate the given technology. All the approaches introduced in this section present intrinsic limitations that are out of scope for this review and are extensively addressed elsewhere.

## CORE CHALLENGES

### *Genetic drift*

Genetic drift is one of the most influencing factors affecting the repeatability of *in vitro* experiments. Instability is associated to clonal dynamics and appearance of new genetic variants. A sample study shows that in 106 cell lines compared across two labs, up to 90% of non-silent mutations were discordant between datasets[124]. MCF-7, specifically, had 27 strains with high genomic diversity, including differential mutations, copy number alterations, and gene expression.[124] These variations are triggered by variations in growth medium inducing clonal shifts, and even single-cell-derived clones are genomically unstable after long-term culture. This has tremendous consequences for drug discovery research. Out of 321 anti-cancer drugs tested across the 27 MCF-7 strains, almost 75% of effective compounds in some strains were completely inactive in others.[124] In a different study, while researching LUCA-15 function in breast cancer, conflicting reports were found regarding its presence in MCF-7 cells, prompting them to characterize their available sublines. It was reported that the chromosomal region where LUCA-15 maps is unstable in MCF-7 cells, and one subline entirely lacked the LUCA-15 gene[13]. This loss correlated with reduced sensitivity to TNF-α–induced apoptosis, which is predominant in MCF-7, since this cell line does not have caspase activity, while overexpression of LUCA-15 restored apoptotic responsiveness. These findings suggest LUCA-15 is highly susceptible to genetic drift, leading to different apoptotic behavior in the same cell line. Although most of the studies refer to MCF-7 as the most unstable breast cancer cell line, it has been demonstrated that HCC1143, HCC38, HCC1937, T47D, BT-549 and MDA-MB-361 are also highly unstable[17]. MDA-MB-231 and HCC1806 are also moderately unstable, while control cell lines such as MCF710A are usually stable. Approaches to minimize these drawbacks include utilizing cell lines at low passages and standardizing culture conditions for all experimental settings. Genetic drift poses challenges in unifying genomic databases and topic-trained AI assistants, while at the same time making their potential impact even more consistent.

### *Conflicting characterizations*

Key information such as age, grade, tumor size, ER/HER2 status, and race is also missing or inconsistently reported across many studies, limiting the utility of the data for epidemiological or prognostic modeling. A good example of these challenges in harmonizing data from different breast cancer datasets is observed in **Table 3**, which compares three of the most commonly used breast cancer databases.[120] The first issue revolves around data collection: CCLE and TCGA-BRCA contain RNA-Seq data, METABRIC contains microarray data. RNA-Seq is much more sensitive than microarrays, leading to technical bias when bringing data together.[125] Furthermore, as discussed earlier, the drug response and characteristics of cell lines are often non-comparable with those of primary tumors due to genetic drift and the absence of a relevant microenvironment. Because of this, it is challenging to associate cell-line data from CCLE to equivalent tumors from TCGA-BRCA or METABRIC and vice versa. While each database provides valuable insights, we cannot directly link cancer cell line features to population-based outcomes without accounting for the underlying biases that affect their representativeness. A multi-omic comparison of 57 breast cancer cell lines to 1019 metastatic breast cancer patient samples from METABRIC and TCGA revealed a mismatch between commonly used cell lines and the genomic landscape of metastatic tumors.[126] One of the most critical findings was that many of the breast cancer cell lines used to model metastatic disease, such as MDA-MB-231, showed poor genomic similarity to actual metastatic breast cancer tumors, particularly within the basal-like subtype. Through a large-scale integrative analysis comparing 57 cell lines with over 1000 metastatic tumor samples, the authors demonstrate that historical model selection often lacks molecular justification. In contrast, less frequently used lines like HCC38, HCC1395, and BT-549 were found to more closely resemble the genomic and transcriptomic profiles of metastatic tumors.[126]

**Table 3. The three biggest and most widely used breast cancer databases in research.** Cancer Cell Line Encyclopedia (CCLE), Cancer Genome Atlas Breast Invasive Carcinoma (TCGA-BRCA) and METABRIC - EGA European Genome-Phenome Archive.

| Feature | CCLE | TCGA-BRCA | METABRIC |
|---|---|---|---|
| Sample type | Breast cancer cell lines | Primary tumor | Primary tumor |
| Sample size | 76 | 1098 | 1992 |
| Expression platform | RNA-seq | RNA-seq | Microarray |
| Subtype distribution | Skewed toward TNBC, HER2+ | Balanced; ER+ enriched | ER+ enriched |
| Drug response | Yes | No | No |
| Passage information | Limited | NA | NA |
| Tissue source | *In vitro* cultured cells (varied origins) | Fresh frozen tissue | Fresh frozen tissue |
| Ancestry diversity | Limited/ not annotated | Mostly European ancestry | UK/Canadian, white |
| Population metadata | Limited | Rich | Moderate |

### *Underrepresentation*

The underrepresentation of cell lines from younger patients and non-Caucasian ethnicities is another limitation of most breast cancer models, affecting the translational relevance particularly for studies aimed at investigating racial disparities in TNBC outcomes.[118] Healthcare access barriers lead to late-stage diagnoses in Hispanic and Asian women, contributing to fewer available data and lower reported incidence. In developed countries, earlier detection is supported by better screening access, while socioeconomic disparities, poor cancer literacy, unhealthy diets, and obesity contribute to higher breast cancer mortality in certain ethnic groups. This disparity continues to bias clinical trials and treatments toward Caucasian patients, further exacerbating healthcare inequalities and limiting the effectiveness of treatments for other ethnic groups. Moreover, most available breast cancer cell lines are derived from older patients, with relatively few from those under 40. Breast cancer from a 23-year-old can be substantially different from the same type of cancer in a 73-year-old. These differences impact how cancer cells behave and respond to treatment, making age another crucial factor to consider. This is becoming increasingly critical because many studies highlight the rising incidence of breast cancer in younger individuals, yet very few of them employ cell lines derived from younger patients (e.g., MCF-7 is from a 69-year-old woman). Integrating all features of every commercial cell line in a common interactive platform would help identifying gaps, misclassifications, and inform a more reliable and relevant experimental design.

## CONCLUSION

This work highlights the need for an integrated multifactorial comparison for the vast and heterogeneous dataset of commercially available cell lines, their derived advanced and emerging models, and their effective clinical relevance, with the aim of improving translational chances of *in vitro* cancer research. It focuses on breast cancer but could be translated to any other cancers, as well as to a set of other diseases commonly studied in bioengineering through the described approaches. It proposes a starting framework to conceive and integrate cutting-edge models that could benefit from contamination straight from the experimental design stage. The ability to align laboratory research with the clinical, molecular, and functional complexity of tumors observed in patients is of crucial importance. The proposed framework suggests initiating with cell lines, the simplest available model, in the context of breast cancer, a well-studied disease, by structuring cell line selection around their absolute (e.g., origin, receptor status, mutations) and relative (e.g., metastatic potential, drug resistance, immune interaction) characteristics to enable a more rational and grounded use of existing resources. The approach, despite statically presented, carries an intrinsically dynamic nature given by the possibility of integrating the proposed blocks into AI-assisted bioinformatic resources. Interfacing the latter with emerging disease models such as 3D organoids and co-culture systems would enable a bidirectional information flow between offices, laboratories, and clinics. In parallel, new ideas worth exploring in the biotechnology front for further integration include cancer virtual modeling using digital avatars trained on multi-omics and functional assay data, automated scoring systems for culture scalability and model fitness, and real-time feedback platforms that dynamically adjust cell line recommendations based on evolving biological parameters. These paradigms point to a future where cell model selection turns into a strategic and adaptive process.

## ACKNOWLEDGMENTS

This work was supported by National funds from FCT - Fundação para a Ciência e a Tecnologia, I.P., with dedicated funds from the project eOnco (2022.07252.PTDC) and institutional funds from iBB (UIDB/04565/2020 and UIDP/04565/2020), and the Associate Laboratory i4HB (LA/P/0140/2020). The authors also received financial support from “la Caixa” Foundation (ID 100010434) LCF/BQ/PI22/11910025. This work is funded by national funds through FCT, I.P., under UID/00481 – TEMA and the project CarboNCT (2022.03596.PTDC)

## DECLARATION OF INTERESTS

The authors declare no competing interests.